\documentclass[aps,prd,nofootinbib,twocolumn,superscriptaddress,floatfix]{revtex4}
 
\usepackage{amsmath,amssymb}
\usepackage[dvips, pdftex]{graphicx}
\usepackage[tight,footnotesize]{subfigure}
\usepackage[usenames]{color}
\usepackage[dvipsnames]{xcolor}
\usepackage{float}
\usepackage{wrapfig}
\usepackage{hyperref}
\usepackage{mathtools}
\usepackage[utf8]{inputenc}
\usepackage{graphicx}
\usepackage{physics}
\graphicspath{ {./images/} }

\usepackage{xcolor}

\begin{document}

\newcommand{\tcb}{\textcolor{blue}}

\def\ga{\mathrel{\raise.3ex\hbox{$>$\kern-.75em\lower1ex\hbox{$\sim$}}}}
\def\la{\mathrel{\raise.3ex\hbox{$<$\kern-.75em\lower1ex\hbox{$\sim$}}}}

\def\be{\begin{equation}}
\def\ee{\end{equation}}
\def\bea{\begin{eqnarray}}
\def\eea{\end{eqnarray}}

\def\betap{\tilde\beta}
\def\del{\delta_{\rm PBH}^{\rm local}}
\def\Msun{M_\odot}
\def\Rcl{R_{\rm clust}}
\def\fPBH{f_{\rm PBH}}

\newcommand{\Mpl}{m_{\rm p}} 
\newcommand{\mpl}{m_\mathrm{pl}} 



\title{Were you born in an aborted primordial black hole?}

\author{Emilie Despontin}

\affiliation{Service de Physique Th\'eorique, Universit\'e Libre de Bruxelles (ULB), Boulevard du Triomphe, CP225, 1050 Brussels, Belgium.}

\author{Sebastien Clesse}

\affiliation{Service de Physique Th\'eorique, Universit\'e Libre de Bruxelles (ULB), Boulevard du Triomphe, CP225, 1050 Brussels, Belgium.}

\author{Albert Escriv\`a}
\affiliation{Division of Particle and Astrophysical Science, Graduate School of Science, \\ Nagoya University. Nagoya 464-8602, Japan}

\author{Cristian Joana}
\affiliation{CAS Key Laboratory of Theoretical Physics, Institute of Theoretical Physics, Chinese Academy of Sciences, Beijing 100190, China}
\affiliation{Cosmology, Universe and Relativity at Louvain (CURL), %
Institut de Recherche en Mathematique et Physique, %
University of Louvain, 2 Chemin du Cyclotron, 1348 Louvain-la-Neuve, Belgium}

\begin{abstract}
We propose a mechanism of electroweak baryogenesis based on the Standard Model and explaining the coincidence between the baryon and Dark Matter (DM) densities.   Large curvature fluctuations slightly below the threshold for Primordial Black Hole (PBH) formation locally reheat the plasma above the sphaleron barrier when they collapse gravitationally, leading to regions with a maximal baryogenesis at the Quantum Chromodynamics epoch.  Using numerical relativity simulations, we calculate the overdensity threshold for baryogenesis.  If PBH significantly contribute to the DM, aborted PBHs can generate a baryon density and an averaged baryon-to-photon ratio consistent with observations.
\end{abstract}
\maketitle
\setlength\parindent{0pt}

\textbf{Introduction --} Despite great efforts over the last fifty years, the nature of Dark Matter (DM), constituting about 85\% of the matter in the Universe, still remains one of the deepest mysteries in Physics. Another equally important enigma concerns the remaining 15\%, which is made of ordinary matter, because the Standard Model (SM) of particle physics alone cannot generate the observed matter-antimatter asymmetry and the associated baryon-to-photon ratio. 
Moreover, it is rather intriguing that DM and ordinary matter have comparable abundances even though they come from radically different production mechanisms.\\  

One possible DM candidate showing a booming interest since 2015, with the detection of gravitational waves from black hole binary coalescences~\cite{LIGOScientific:2016aoc}, is the case of primordial black holes (PBHs). Large primordial inhomogeneities may have collapsed into PBHs when they re-entered the cosmic horizon, if they were exceeding some overdensity threshold. PBH formation is typically boosted at the Quantum-Chromodynamics (QCD) epoch due to the transient reduction of this threshold, leading to a strong peak in the abundance of stellar-mass PBHs~\cite{Byrnes:2018clq,Carr:2019kxo}. If they significantly contribute to DM, it was pointed out in ~\cite{Carr:2019hud,Garcia-Bellido:2019vlf} that their abundance at formation, denoted $\beta_{\rm PBH}$, is comparable to the baryon-to-photon ratio $\eta^{\rm tot} \simeq 6 \times 10^{-10}$. This may suggest a connection between the baryogenesis mechanism at the origin of ordinary matter and DM in the form of solar-mass PBHs from the QCD epoch. This possibility is constrained by several probes, see e.g. \cite{Carr:2020xqk,Carr:2020gox,Escriva:2022duf,LISACosmologyWorkingGroup:2023njw} for reviews on PBHs and the limits on their abundance, but could at the same time explain various cosmic conundra, as advocated in \cite{Clesse:2017bsw,Carr:2019kxo,Carr:2023tpt}, especially if PBHs have a wide mass distribution.

A connection between baryogenesis and PBHs was considered in several theoretical frameworks~\cite{Barrow:1990he,Upadhyay:1999vk,Rangarajan:1999zp,Dolgov:2000ht,Baumann:2007yr,Dolgov:2008wu,Hook:2014mla,Aliferis:2014ofa,Hamada:2016jnq,Morrison:2018xla,Carr:2019hud,Garcia-Bellido:2019vlf,Aliferis:2020dxr,Hooper:2020otu,ShamsEsHaghi:2022azq,Gehrman:2022imk}, such as through the evaporation of tiny PBHs by Hawking radiation~\cite{Baumann:2007yr,Dolgov:2000ht,Hook:2014mla,Morrison:2018xla,Hooper:2020otu,Boudon:2020qpo,Perez-Gonzalez:2020vnz,Datta:2020bht,Gehrman:2022imk,Bernal:2022pue}. In particular, evaporating PBHs may have locally reheated the primordial plasma enough to induce electroweak baryogenesis at low temperature~\cite{Upadhyay:1999vk,Rangarajan:1999zp,Aliferis:2014ofa,Aliferis:2020dxr}. 
Local reheating may have also been induced by other processes, as in~\cite{Asaka_2004,FireballBaryo}. 
Alternatively, Carr, Clesse and Garc\'ia-Bellido (CCGB) proposed in~\cite{Carr:2019hud,Garcia-Bellido:2019vlf} a mechanism where the region around forming PBHs at the QCD epoch is reheated due to gravitational collapse.

In this \textit{letter}, we explore a new scenario 
where the inhomogeneities slightly below the threshold value collapse and reheat the plasma locally, but for which the pressure prevents PBH formation.  We use numerical relativity simulations in order to study such aborted PBHs and determine the maximum density and temperature that can be reached. Then, we show that the conditions for electroweak baryogenesis are met and we calculate the local and averaged baryon-to-photon ratio.  Our calculations rely on the formalism of~\cite{Carr:2019hud,Garcia-Bellido:2019vlf}, with several improvements and corrections. 
It is worth noticing that in our scenario, the CP violation is only provided by the SM and 
baryogenesis is achieved through the only known interaction not included in the SM, gravitation. This scenario naturally connects the baryon and PBH abundances
as well as the initial PBH abundance to the baryon-to-photon ratio.\\

\textbf{Baryogenesis from sphaleron production --} Even if the SM alone cannot quantitatively explain the observed matter-antimatter asymmetry, it embeds a mechanism to produce such an asymmetry through the production of so-called \textit{sphalerons}~\cite{GAVELA_1994,Huet:1994jb}.  
They are electroweak gauge field configurations acting like a barrier between different vacuum states. Sphaleron transitions violate the conservation of the baryon number. However, at thermal equilibrium, the conversion of leptons into baryons is compensated by the inverse process. As a consequence, in order to prevent such a washout and obtain a net asymmetry, a process driving the system into non-equilibrium is necessary, such as through a first order phase transition \cite{Khlopov:1985jw,Jedamzik:1999am,Liu:2021svg} or reheated regions by PBH evaporation \cite{Datta:2020bht,Smyth:2021lkn}. In our model, hot spots are produced by the gravitational collapse of relatively large-amplitude curvature fluctuations that remain below the threshold of PBH formation.   
\\

In the early Universe, the photon number density $n_\gamma$ is related to the entropy density \cite{peter2009primordial}
\begin{equation}
    s=\frac{2\pi^2}{45}g_{*\rm s} T_{\mathrm{th}}^3 = \frac{\pi^4}{45}\zeta(3) g_{* \rm s} n_\gamma \simeq 1.8 \, g_{* \rm s} n_\gamma.
\end{equation}
where $g_{* \rm s}$ is the effective number of degrees of freedom associated with the entropy $s$, $T_{\rm th}$ is the thermal temperature of the Universe and  $\zeta$ is the Riemann zeta function. 
The local baryon-to-photon ratio is then given by 
\begin{equation}
    \eta^{\rm loc} \equiv \frac{n_{\rm b}}{n_\gamma} \simeq \frac{1.8 \, g_{* \rm s} \,n_{\rm b}}{s} =  \frac{1.8 \cdot 2 \pi^2 n_{\rm b}} {45\, T_{\mathrm{th}}^3 } .
\end{equation} 




The baryon number density $n_{\rm b}$ in one of those hot spots follows an approximate Boltzmann equation \cite{Carr:2019hud}
\begin{equation} \label{eq:boltzmann}
	\dv{n_{\rm b}}{t}+\Gamma_{\rm b} n_{\rm b} = \Gamma_\mathrm{sph}\frac{\mu_\mathrm{eff}}{T_{\rm r}},
\end{equation}
where $\Gamma_{\rm b}=(39/2)\,\Gamma_\mathrm{sph}(T)/T^3$ and $\mu_\mathrm{eff}=\delta_{\mathrm{CP}}(T)\, \text{d}\theta/\text{d}t$ is the effective chemical potential for the baryon production.  
$T_{\rm r}$ is the temperature reached in the reheated region.  For the SM CP violation parameter, 
we rely on~\cite{GAVELA_1994} and assume%
\footnote{
It is important to note the difference with~\cite{Garcia-Bellido:2019vlf,Carr:2019hud} which assumed that  $\delta_{\rm CP} \propto T^{-12}$ based on ~\cite{shap_1993}. However, this relation was refuted in ~\cite{GAVELA_1994}. When using Eq.~\eqref{eq:deltaCP}, we find that the mechanism of~\cite{Garcia-Bellido:2019vlf} is not efficient enough to explain the observed value of $\eta$.
}

\begin{equation} \label{eq:deltaCP}
		\delta_{\mathrm{CP}}(T) = 1.6 \cdot 10^{-21}\, \frac{T}{\rm GeV}.
\end{equation}

The sphaleron transition rate $\Gamma_{\rm sph}$ strongly depends on the temperature and is given by~\cite{Carr:2019hud,D_Onofrio_2014,hong2023baryogenesis}
	\begin{equation} \label{eq:Gammasph}
		\Gamma_\mathrm{sph} (T) \simeq \begin{cases}
				30 \,\alpha_W^5 T^4 & \text{if }T>T_{\rm c}, \\
				\left(\frac{E_\mathrm{sph}}{T}\right)^3 m_{\rm W}^4(T) e^{-\frac{E_\mathrm{sph}}{T}} & \text{if }T<T_{\rm c}, \\
			\end{cases}
	\end{equation}
where $T_{\rm c} \simeq 150$ GeV is a critical temperature below which the sphaleron production rate is exponentially suppressed. In Eq.~\eqref{eq:Gammasph}, $\alpha_{\rm W} = 1/29$ is the weak coupling constant, the sphaleron effective energy is ${E_\mathrm{sph} \simeq 2m_{\rm W}/\alpha_{\rm W}}$ and the squared of the electroweak mass is ${m_{\rm W}^2(T)=\pi \alpha_{\rm W}\left[ v_0^2\left(1-T^2 / 12 v_0^2\right)  ^2 +T^2\right]}$ where ${v_0 = 245\ {\rm GeV}}$  is the Brout-Englert-Higgs expectation value at zero temperature.
\\

Because 
${T_{\rm th} \sim 100 \,{\rm MeV} \ll T_{\rm c} < T_{\rm r} \sim {\rm TeV}}$, sphaleron transitions are immediately extinguished when highly energetic baryons exit the reheated bubble. One can thus neglect the second term of Eq.~\eqref{eq:boltzmann}. Integrating over time, the gain in baryon number  of $\Delta B = 3 \Delta N_{\rm cs}$ where $N_{\rm cs}$ are the Chern-Simons numbers, is obtained after a time lapse necessary for a Higgs phase variation of $\Delta \theta \sim \pi$.  Most baryons produced through the sphaleron process in the reheated regions hit the cold surrounding environment, such that the generated baryon number density is frozen and can be approximated by 
\begin{equation}
    n_{\rm b} \simeq \pi \,\Gamma_\mathrm{sph}(T_{\rm r})\frac{\delta_{\mathrm{CP}}(T_{\rm th})}{T_{\rm r}}~.
\end{equation}
\\
One can therefore estimate the baryon asymmetry produced during the collapse if one knows the  temperature that is reached during the local reheating.\\

\textbf{Reheating temperature --}  Let us now consider the process for generating hot regions 
during the QCD transition and proton freeze-out (${20 \,\text{MeV} < T< 200 \, \text{MeV}}$). The effective temperature reached inside the collapsing inhomogeneity is related to the kinetic energy $E_{\rm k}$ acquired by the protons accelerated by the gravitational collapse, $T_{\rm r} = \frac 2 3 E_{\rm k}$. Initially, the inhomogeneity has a super-Hubble size and is expanding. It re-enters the Hubble radius $d_{\rm H}$ at time $t_*$ and starts contracting until it reaches a minimal radius, parameterized as $\gamma \, d_{\rm H}(t_*)$, when the pressure takes over the gravitational force. The available energy $\Delta E$ is related to the change in the gravitational potential during the collapse,
\begin{align}
		\Delta E 
		& \simeq \frac{M_{\rm H}^2}{d_{\rm H}}\left(-1+\frac1\gamma\right),
\end{align}
with the Hubble mass $M_{\rm H}=d_{\rm H} {c^2}/{G} $, or in terms of the thermal temperature at horizon crossing $T_*$,
\begin{equation}
	M_{\mathrm{H}}\simeq 0.5\, g_*^{-1/2}M_\odot\left(\frac{T_*}{\text{GeV}}\right)^{-2}.
\end{equation}
This energy reservoir is shared between species, so at the QCD transition one has
\begin{equation}
 E_k = \frac{\Delta E \left({n_{\rm p}}/{n_\gamma} \right) (1/ n_{\rm p} ) }{(4 \pi/3)\,   \gamma^3 d_H^3 }  
\end{equation}
where $n_{\rm p}$ is the number density of protons.  A similar temperature is obtained if one considers the acceleration of quarks before the QCD transition. 
The smaller the value of $\gamma$, the larger the kinetic energy and the higher the reheating temperature, which can eventually go above the sphaleron barrier, $T_{\rm r} \gtrsim 150 {\rm GeV}$.   
For the proton number density, we use the ideal gas approximation. In thermal equilibrium, it follows a Fermi-Dirac distribution with 2 spin states. Below 200 MeV, protons are non-relativistic and non-degenerate and we can use the Maxwell-Boltzmann approximation to write the number density as
 \begin{align}
		n_{\rm p}  &=  2\left(\frac{2\pi m_{\rm p} T_*}{h^2}\right)^{3/2} { e}^{- m_{\rm p}/T_*} \nonumber  \\
  & \simeq  1.37 \cdot 10^{46}  \left(\frac{m_{\rm p}}{T_*}\right)^{-3/2}e^{-m_{\rm p}/T_*} \left[\rm m^{-3} \right],
\end{align}
 where $m_{\rm p} $ denotes the proton mass.
Therefore, the kinetic energy acquired by the protons from the collapse is about 
\begin{equation}
    E_{\rm k}  \simeq 4.63\, g_*(T_*)\frac{1-\gamma}{\gamma^4}\left(\frac{m_{\rm p}}{T_*}\right)^{-5/2}e^{m_{\rm p}/T_*} \,\text{GeV}~.
\end{equation}

\begin{figure}[t!]
     \centering
     \includegraphics[width=8.5cm]{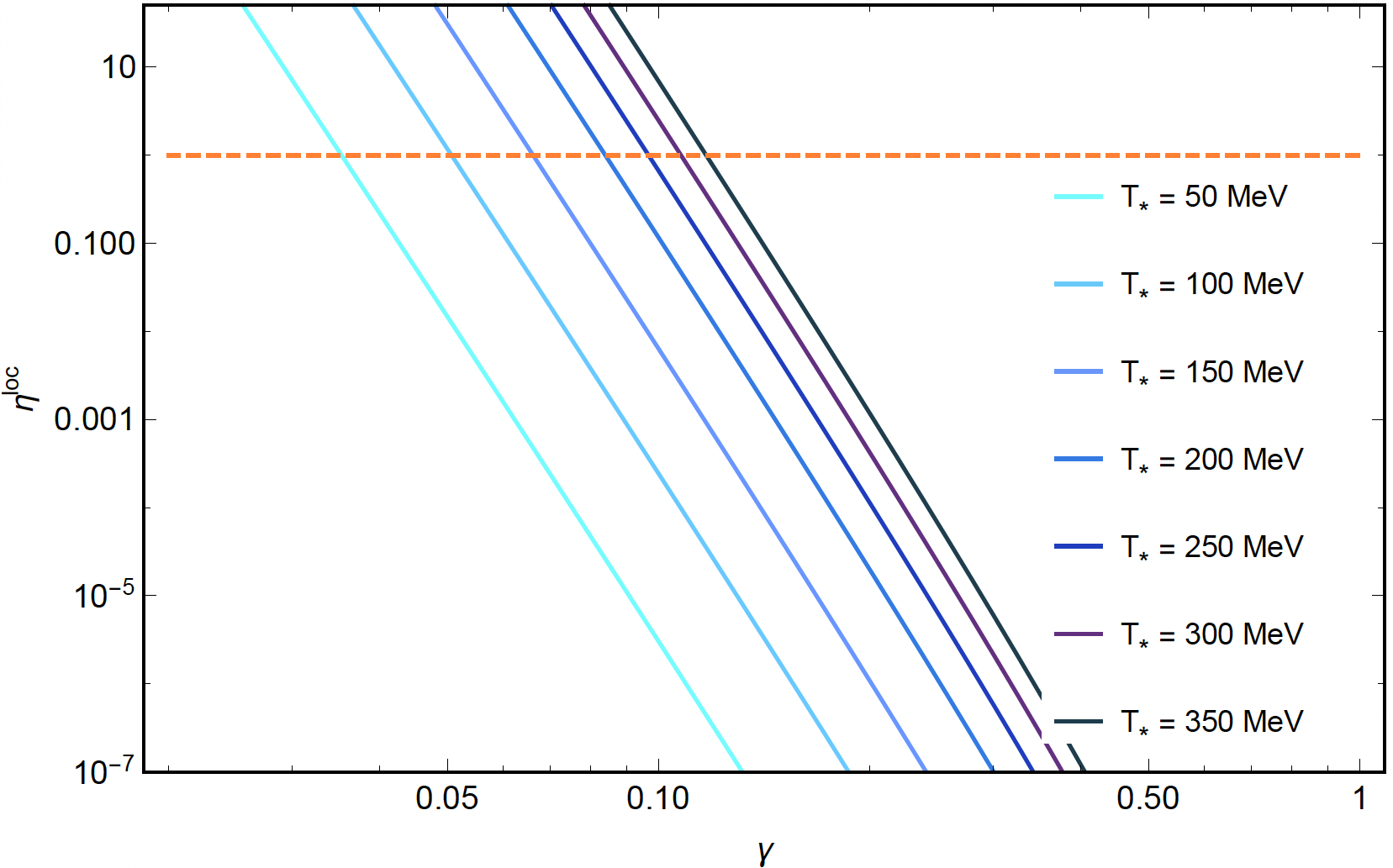}
    \caption{%
      Expected local baryon-to-photon ratio as a function of the compression factor $\gamma$ and the background thermal temperature. 
      \label{fig:etaSVgamma}
      }
\end{figure}

Before investigating the realistic values of $\gamma$ that can be reached during the collapse, we have computed in Fig.~\ref{fig:etaSVgamma} the value of gamma required to obtain a local baryon-to-photon ratio $\eta^{\rm loc} \ge 1$ from the collapse. Because the process is not instantaneous, one cannot use $T_{\rm th} = T_*$. Based on numerical simulations described thereafter, we have estimated that $T_{\rm th} \simeq T_* / 4.8$, but small changes do not influence the validity of the scenario.\\

\begin{figure*}[t!]
    \centering
    \includegraphics[width=8.5cm]{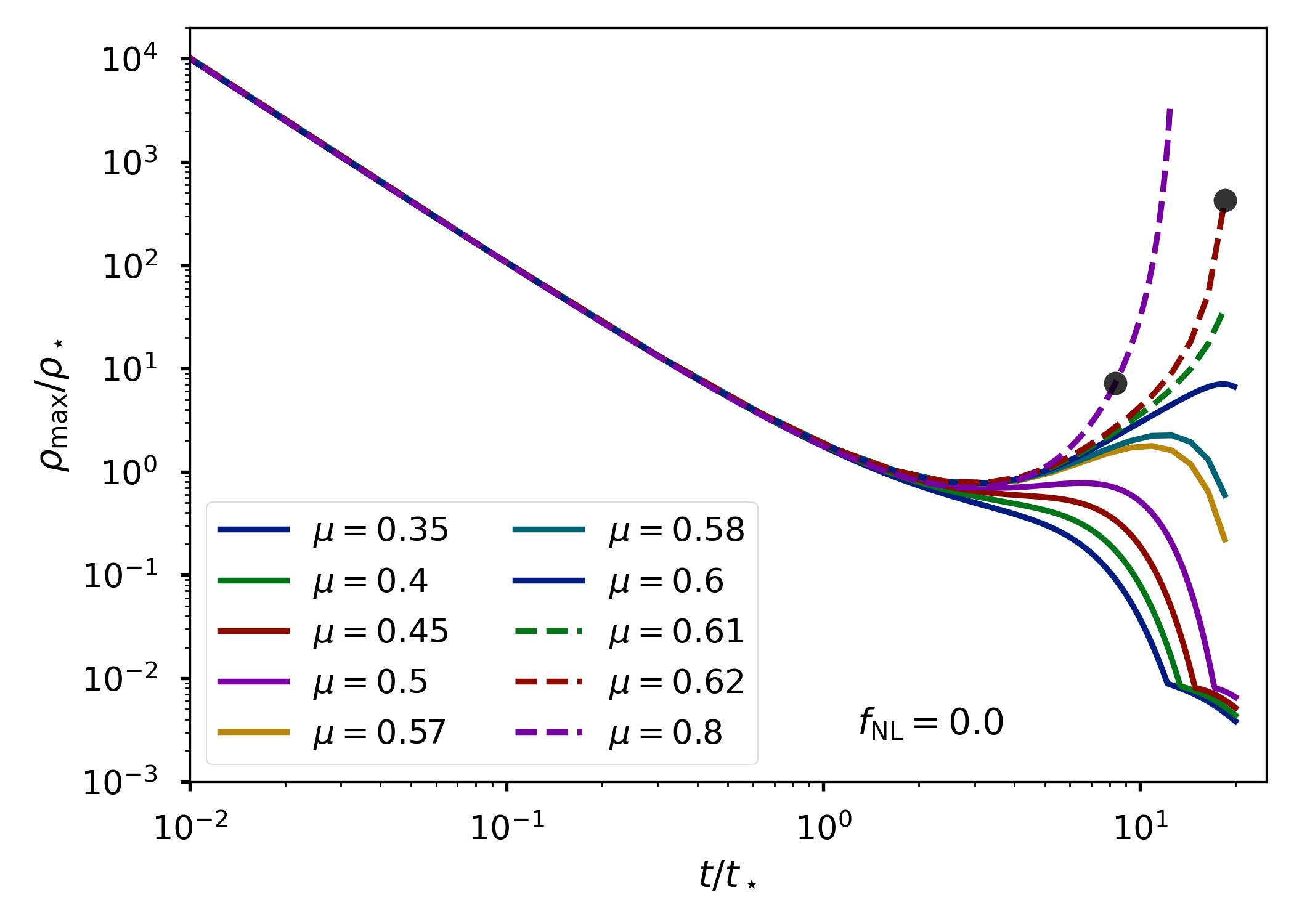}
    \includegraphics[width=8.5cm]{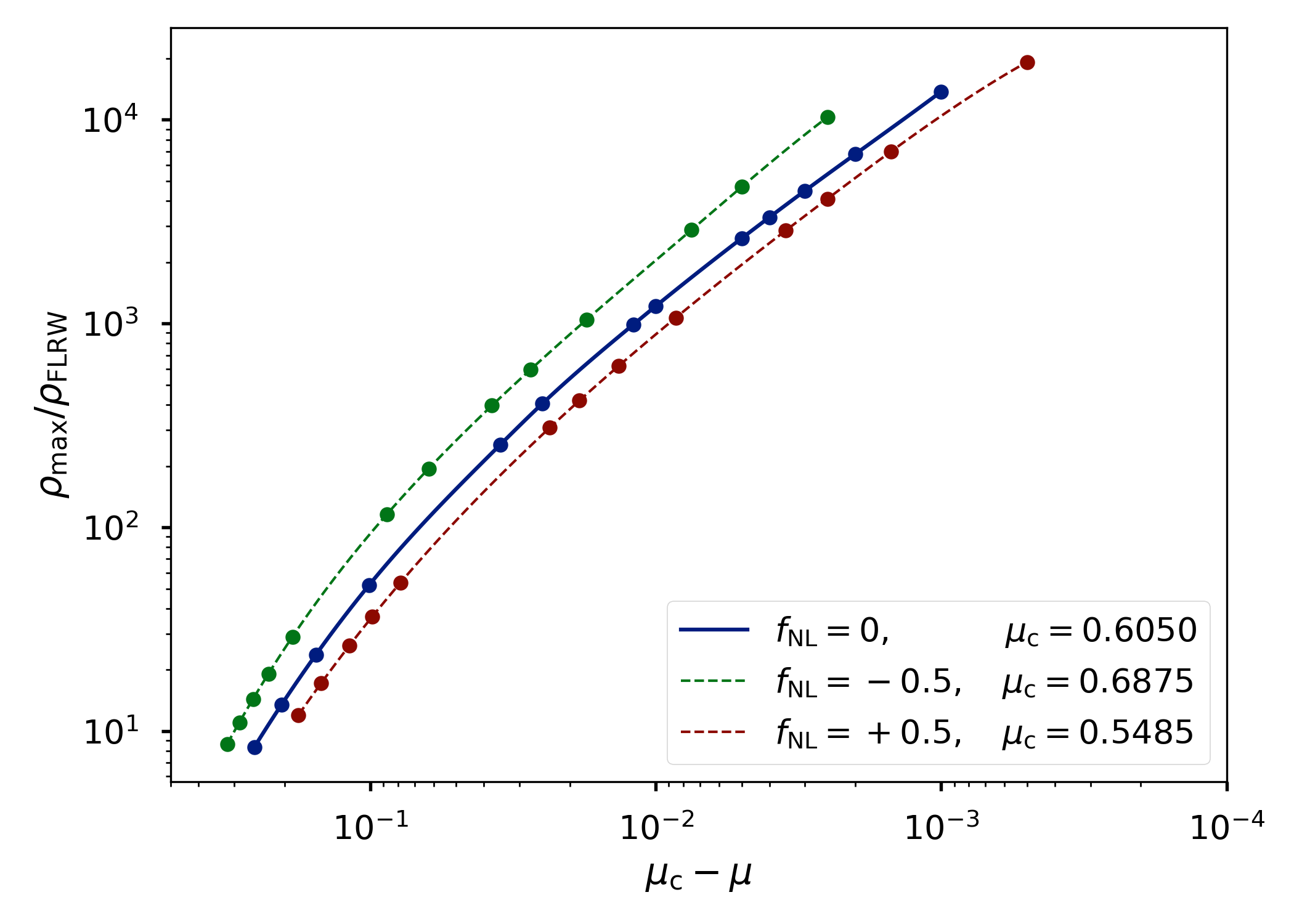} 
    \caption{Left: time-evolution of the maximal density for several amplitudes of curvature perturbations, with $f_{\rm NL}=0$, where $\delta=(4/9) \mu$ at horizon entry $t_*$.  
    Solid lines correspond to the cases with \textit{aborted} collapse where no PBH is formed. Dashed lines correspond to supra-threshold amplitudes and the timing for black hole formation is labeled with black dots. 
    Right: ratio of maximal energy density over FLRW background energy density
    for perturbations at sub-threshold amplitude difference $(\mu_{\rm c} - \mu)$ and various values for $f_{\rm NL} = -0.5,\ 0,\ 0.5$ labeled with green, blue and red lines, respectively.  }
    \label{fig:pics_density}
\end{figure*}

\textbf{Simulations of aborted PBHs -- } %
We have run several numerical relativity simulations by solving the Misner-Sharp equations in the comoving gauge \cite{Misner:1964je} under the assumption of spherical symmetry. 
As in ~\cite{Polnarev:2006aa,Musco:2018rwt,Escriva:2021aeh}, we use the gradient expansion formalism at second order to create the initial data with the long-wavelength approximation approach. The simulation grid is discretized along the comoving radial dimension $r$ and spatial derivatives are computed either with fine-element stencils at fourth order or by employing pseudo-spectral methods as in \cite{Escriva:2019nsa}. 
With this formulation, the areal radius is given by $R = a(t) r \exp[{\cal R}(r)] $, where $a(t)$ is the background scale factor and ${\cal R}(r)$ is the scalar curvature perturbation in the comoving gauge. We use a quadratic expansion for the profile of the curvature perturbation,
\begin{equation}
{\cal R}(r) = {\cal R}_{\rm G}(r)  + \frac 35 f_{\rm NL}  {\cal R}_{\rm G}^2(r)  ~,
\end{equation}
where we chose the Gaussian profile ${\cal R}_{\rm G}(r) \equiv \mu\, \mathrm{sinc} \left( k_\star r \right)$ where 
the strength of non-Gaussianity is parameterized by $f_{\rm NL}$ and $k_\star$ is the characteristic scale of the perturbation.  
We have considered three different values of $f_{\rm NL}$: vanishing, positive or negative with $f_{\rm NL} = \pm 0.5$ in order to illustrate the possible effect of non-Gaussianities.
\ \\

We use these numerical simulations to investigate the formation of overdensities or \textit{hot spots} due to the gravitational pull after the Horizon re-entry of the curvature perturbation and we establish the threshold in the curvature fluctuation amplitude above which baryogenesis is expected to take place. In Fig.~\ref{fig:pics_density} (left panel), we display the maximal density along the radial profile as a function of time, for different amplitudes $\mu$ that can be approximately related to the overdensity by $\mu \approx (9/4) \delta$. The compression factor is then computed as $\gamma (t) = \left(\rho_{\max} (t)/\rho_*  \right) ^{-1/3}$ with $\rho_* \equiv \rho (t_{*})$.
From these results, in the right panel, we have extracted the maximal density value reached during the simulation over the background density at the time of Hubble horizon crossing.  It is used to compute the minimum value of $\gamma$ reached during the collapse. In turn, we obtain the overdensity threshold for baryogenesis, denoted $\delta_{\rm baryo}$, that can produce a low enough $\gamma$ to lead to $\eta^{\rm loc} \geq 1$.
\\

\textbf{Averaged baryon-to-photon ratio -- }%
%
The next step is to compute the abundance of density fluctuations 
 to obtain the averaged baryon-to-photon ratio and examine the connection with the PBH abundance.  
Although there are a lot of PBH formation mechanisms, they have a common feature: large overdensities, 
commonly assumed to be of inflationary origin, above the threshold $\delta_{\rm c}$, which generally depends on the equation of state and density profile.  For a Gaussian probability distribution $P(\delta)$, one obtains a density of PBHs $\rho_{\rm PBH}$ at formation given by
\begin{equation}
\beta(m_{\rm PBH}) \equiv \frac{{\rm d}\,(\rho_{\rm PBH} / \rho_{\rm c})}{{\rm d} \ln m_{\rm PBH}} = \int_{\delta_{\rm c}}^{\infty}P(\delta)\, {\rm d}\delta,
\label{Eq:beta}
\end{equation}
$P(\delta)$ is characterized by the root-mean-squared of density fluctuations $\delta_\mathrm{rms}$, an adjustable parameter related to the primordial power spectrum from inflation.  As an illustrative example, we have considered an almost scale-invariant spectrum ${\delta_{\rm rms} = A \left(M/{M_\odot}\right)^{(1-n_s)/4}}$, with an amplitude $A=0.0897$ and a spectral index $n_s=0.97$, following \cite{Carr:2019kxo}. 
At the same time, we are interested by the averaged baryon-to-photon ratio induced by aborted PBHs of mass $m_{\rm PBH}$, obtained by integrating the probability distribution of density fluctuations from $\delta_\mathrm{baryo} $ to $ \delta_c$,
\begin{equation}
	\eta (T_*) \simeq \int_{\delta_\mathrm{baryo}}^{\delta_c}P(\delta)\,{\rm d}\delta,
 \label{Eq:etaTstar}
\end{equation}
and the difference between $\delta_\mathrm{baryo}(T_*) $ and $\delta_\mathrm{c}$ has been computed from our numerical simulations.  For $\delta_\mathrm{c}$, we have used the values recently obtained in~\cite{Escriva:2022bwe} that include the effect of the QCD cross-over transition that typically boosts the PBH formation, producing a high peak and a small bump in the PBH mass function~\cite{Carr:2019kxo,Byrnes:2018clq}. 

\begin{figure}[t!]
    \centering
    \includegraphics[width=8.5cm]{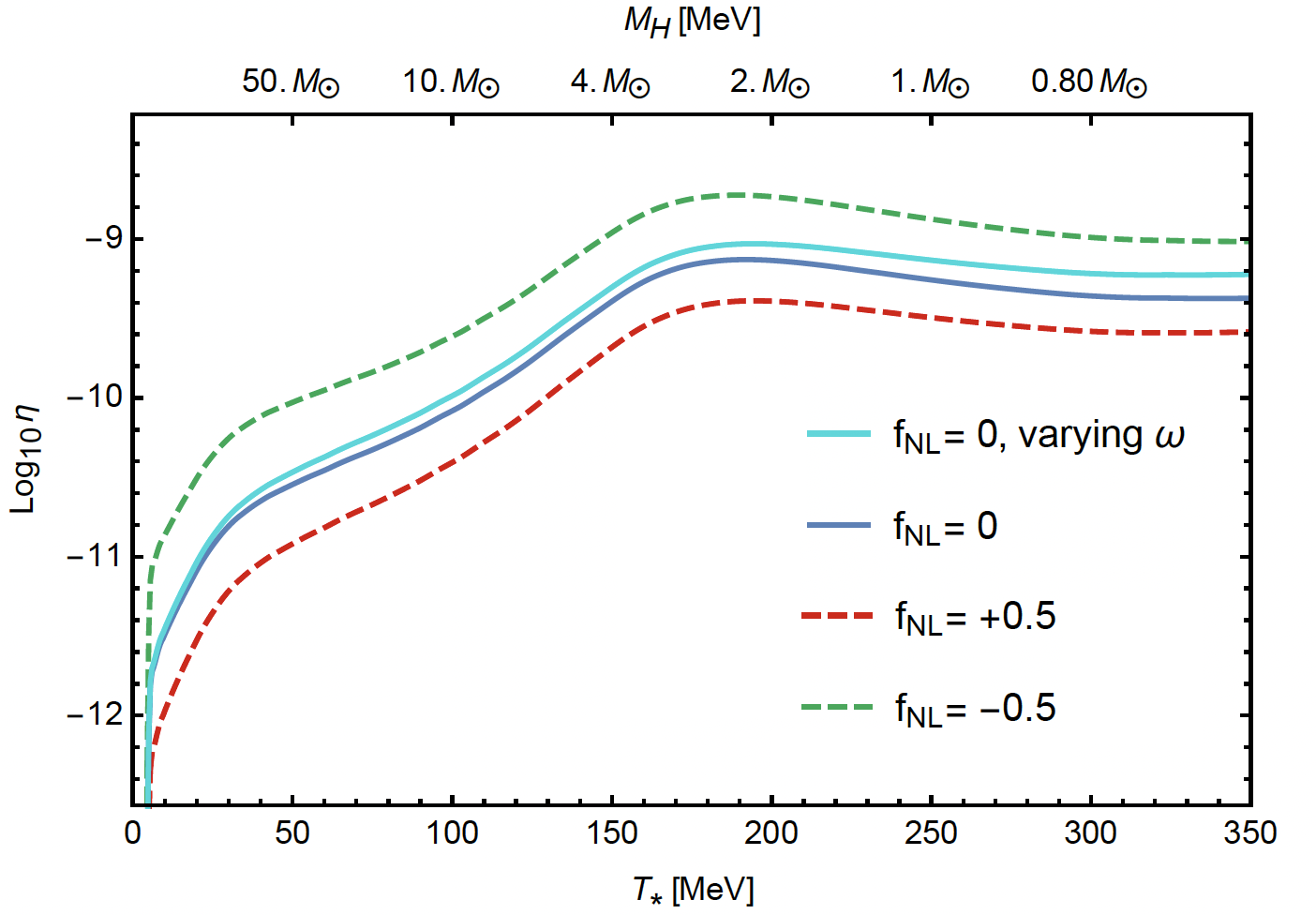}
    \caption{Evolution of the baryon-to-photon ratio $\eta$ produced by the gravitational collapse of fluctuations re-entering in the Hubble radius at temperature $T_*$, for our benchmark illustrative model. 
    The solid blue line is obtained for Gaussian curvature fluctuations ($f_{\rm NL} =0$) and gives an integrated baryon-to-photon ratio of $\eta^{\rm tot} = 6 \times 10^{-10}$ consistent with observations. The dashed red and green lines are obtained for non-Gaussian fluctuations with $f_{\rm NL} = + 0.5$ and $f_{\rm NL} = - 0.5$ respectively, with the same value of the $\delta_{\rm rms}$ parameter. The solid cyan curve is based on our second set of simulations including the variations of the equation of state during the collapse, for Gaussian fluctuations.
    \label{fig:etafinal}
    }
\end{figure}

Fig. \ref{fig:etafinal} shows the total $\eta (T_*)$ that we obtain, which is compatible with the observed baryon asymmetry. The baryogenesis efficiency is maximal between $200$ MeV and $150$ MeV and then decreases at lower temperatures, as it becomes harder to pass the sphaleron barrier, which translates in a threshold value $\delta_{\rm baryo}$ closer to $\delta_{\rm c}$ and therefore a lower value of $\eta$. We applied our calculations above $200$ MeV, where we observe a plateau in $\eta$, however our calculations should be refined in this regime to take into account the various species present before the QCD transition.  
The genericity of our scenario has been tested by considering both Gaussian and non-Gaussian curvature fluctuations,  
but in all cases, the observed baryon asymmetry in the Universe can be accommodated with a slight rescaling of the primordial power spectrum. 

Theoretical uncertainties remain in the value of $\delta_c$ and at the different steps of the presented scenario, nevertheless their effects can be counterbalanced by an adjustment in $\delta_\mathrm{rms}$ in order to get the observed values of $\eta^{\rm tot}$.  One should note, however, that tiny changes in $\delta_\mathrm{rms}$ have a large impact both on $\eta^{\rm tot}$ and $\beta_{\rm PBH}$. But what is remarkable is the natural connection between $\eta^{\rm tot}$ and $\beta_{\rm PBH}$.  Indeed, because $\delta_{\rm c} - \delta_{\rm baryo} \ll \delta_{\rm rms} \simeq 10^{-2}$, Eqs.~\eqref{Eq:beta} and~\eqref{Eq:etaTstar} approximately have the same value. 

Interestingly, this naturally produces a sizable DM fraction in PBHs $f(m_{\rm PBH})$, connected to $\beta$ through~\cite{Carr:2019kxo}
\begin{equation}
    f(m_{\rm PBH}) \simeq 2.4 \, \beta(m_{\rm PBH}) \sqrt{\frac{2.8 \cdot 10^{17} M_\odot}{m_{\rm PBH}}}~,
\end{equation}
for stellar PBH masses.
These PBHs could possibly explain the properties of some black hole binary coalescences seen by LIGO/Virgo/Kagra \cite{Byrnes:2018clq,Juan:2022mir,Clesse:2020ghq}. Due to the expected peak in the PBH mass distribution around $2 M_\odot$, corresponding to $T_* \simeq 140 $ MeV, it is therefore also possible that the whole PBH mass function explains the DM in totality, as recently advocated in~\cite{Carr:2023tpt}. \\ 
 

\textbf{Conclusion and discussion --} 
We have proposed that baryogenesis was produced in aborted PBHs, when the gravitational collapse of curvature fluctuations slightly below the threshold of PBH formation leads to a local reheating, to the production of sphalerons and to a maximal electroweak baryogenesis, without assuming any exotic physics. We have adapted the CCGB mechanism relying on shock waves produced during PBH formation to this idea and combined it with numerical relativity simulations to compute the overdensity threshold for baryogenesis and prove the efficiency of this novel baryogenesis process based on aborted PBHs.  
The baryogenesis is maximal around 150-200 MeV and our model naturally connects the abundance of PBHs formed at this epoch to the baryon-to-photon ratio.  We have also emphasized that the connection with the PBH abundance is generic and linked the small value of the difference between the overdensity thresholds for PBH formation and baryogenesis, typically $\delta_{\rm c} - \delta_{\rm baryo} \lesssim 10^{-2}$.
\\ 



Refined calculations and simulations are still needed to relate the PBH abundance and the associated baryon-to-photon ratio more precisely.   
Uncertainties may also harm the genericity of the mechanism.  Notably, the validity of our assumption that a large fraction of the accelerated protons can exit the reheated spots to reach the colder surrounding regions before the generated asymmetry is washed out, should be investigated cautiously.  
The transfer of potential gravitational energy into hadron kinetic energy is also a complex, non-equilibrium process and one should go beyond order of magnitude estimations with dedicated numerical simulations.   
\\

In summary, we propose a plausible unified explanation of the DM, the matter-antimatter asymmetry and the coincidence problem between baryonic and DM densities, in the context of PBHs at the QCD epoch and without adding any new physics other than the one at the origin of the primordial fluctuations.  The mechanism therefore broadens the interest of PBHs beyond the problems of DM and gravitational waves from black hole mergers and also leads to the philosophically interesting idea that our existence itself may be linked to the formation of black holes in the early Universe.
 \\

\begin{acknowledgments}
We warmly thank Juan Garcia-Bellido, Teruaki Suyama, Shi Pi, Miguel Vanvlasselaer and Sebastian Zell for helpful discussions. This work is supported in part by the National Key Research and Development Program of China Grant No. 2021YFC2203004.  
S.C. and E.D. acknowledge financial support from the Belgian Francqui Foundation through a Start-up grant and the Belgian fund for research FNRS through a MIS grant.  A.E. acknowledges support from the JSPS Postdoctoral Fellowships for Research in Japan (Graduate School of Sciences, Nagoya University). C.J. is funded by the NSFC grant Num. 12347132, he is thankful for the hospitality received while visiting the ULB and acknowledges the usage of CURL cosmo clusters at the University of Louvain, funded by the “Fonds de la Recherche Scientifique FNRS” under Grant Num. T.0198.19. 
\end{acknowledgments}

\bibliography{PBH_biblio.bib}
\end{document}